# COMMUNICATION IS THE UNIVERSAL SOLVENT: ATREYA BOT - AN INTERACTIVE BOT FOR CHEMICAL SCIENTISTS

Mahak Sharma, Vidhya Bhawan Gandhiyan Institute of Educational Studies, India, mahak10sharma@gmail.com

Abhishek Kaushik, Dublin City University; Ireland, abhishek.kaushik2@mail.dcu.ie

Rajesh Kumar, Dublin Business School, Dublin, Ireland, mayanrajesh@gmail.com

Sushant Kumar Rai, Dublin Business School, sush20041993@gmail.com

Harshada Hanumant Desai, Dublin Business School, harshadadesai007@gmail.com

Sargam Yadav, Dublin Business School, sargamyadav719@gmail.com

**Abstract:** Conversational agents are a recent trend in human-computer interaction, deployed in multi-disciplinary applications to assist the users. In this paper, we introduce "Atreya", an interactive bot for chemistry enthusiasts, researchers, and students to study the ChEMBL database. Atreya is hosted by Telegram, a popular cloud-based instant messaging application. This user-friendly bot queries the ChEMBL database, retrieves the drug details for a particular disease, targets associated with that drug, etc. This paper explores the potential of using a conversational agent to assist chemistry students and chemical scientist in complex information seeking process.

## 1. INTRODUCTION

Human-computer interaction is a recent trend in computing research (Gwizdka, 2010; Kaushik et al., 2020a; Radlinski & Craswell, 2017) . Chatbots and voice assistants have become an integrated part of many people's daily life. Numerous studies have been conducted to investigate the usability of chatbots in the current era (Arguello et al., 2018; Avula et al., 2018; Avula & Arguello, 2020; Kaushik et al., 2020b)**.** These studies outline the opportunities and challenges for using chatbots as human-computer interaction application in multiple diverse areas such as ticket booking, flight booking etc., to assist users. These studies are limited to broad applications, whereas still, there is an enormous scope for application of chatbots in specific areas (Araujo, 2020; Kaushik & Jones, 2021)**.** In this current study, we introduce Atreya Bot. This interactive conversational agent enables interactions between humans and machines via a web interface or mobile interface. This agent is specially designed for our chemical scientist to support them in information seeking. Our research investigates the opportunity to reduce the cognitive load in the information-seeking process, which may assist the users in better, effective and efficient information seeking. Atreya is a unique type of bot with many features that focus on sufficing the chemical needs of scientists and researchers. It works on retrieving the following relationships from the ChEMBL databases (Davies et al., 2015; Mendez et al., 2019).: Target-Drugs, Drug-Target, Drug-Disease, Drug-Approval Phase, Drug from gene name, from synonyms SMILES (simplified molecular-input line-entry system). These and many other features (which are discussed in detail further in the paper) make ChEMBL a unique virtual assistant.

The paper is structured to highlight the motivation of the study, an overview of the technical components of Atreya, conclusion and the future scope for this study.

## 2. MOTIVATION

The motivation behind the project is explained into the following sub-sections.

### 2.1 Single query conventional search

A query in the search system returns the most relevant document as per their ranking algorithm in a single shot in a conventional search system. The output obtained is not refined concerning the user's information need. The user may have to make multiple search queries by transforming the search query to meet the information need. There are many limitations associated with a single query search to affect its search success. The following points are the constraints of single-query search, which are intended for application of an interactive search bot (Kaushik, 2019; Kaushik et al., 2020a):

1. User must thoroughly explain the information need in a single query: The user should include all required information in a single query to obtain the most suitable relevant results. On reasonable grounds, this can be a very challenging and difficult task.
2. User may not be able to communicate their information need satisfactorily: Usually, users are not conscious of how to describe the information need in a single query. In common, the user attempts to formulate the search keyword, including words incorporated with the information need. But this may or may not indicate the exact specification of the users.
3. High cognitive load on the user in composing a search query: The point as mentioned above 1 & 2 leads to a cognitive load on the user to form the correct query to obtain the most relevant results meet the user the information need.
4. A search system should return relevant content in a single query: Presenting the best relevant results in a single query is usually a complex task for the search system.
5. The user must inspect returned content to find relevant information: A search engine presents the user with the best results but checking the search result's relevancy is the user's liability, which adds to their cognitive load.

The data in ChEMBL database is simple and structured. Therefore, a simple rule-based approach is used to query the database. The Atreya bot is an IR system that uses domain specific and factual knowledge which can be accessed through simple queries. The bot provides a self-driven approach for the user that eliminates the need to perform complex queries. A conversational support funnel is created to help the user to conduct queries and avoid ambiguity. A custom-made toolkit that utilizes text trimming for user-query extraction. We are exploring the need to perform semantic searches, as the information is objective and fact-based.

Atreya Bot is convenient in terms of usage as compared to the search tool of the ChEMBL database. User cognitive load is shown to be very high when querying on interfaces like the search tool discussed above. Conversational agents have proven to greatly reduce the cognitive load on the user. User Engagement is also shown to increase when using a conversational agent as compared to an traditional search interface as discussed above.

## 3. TECHNICAL OVERVIEW

The section explains the components, operations, workflow and potential functionality and use cases.

### 3.1. Components

Bot developed by using four components: AIML, ChEMBL, Telegram and Cairosvg.

1.     **AIML (Artificial Intelligence Markup Language):** AIML is an Extensible Markup Language (XML) file used to create conversational patterns based on which the conversational agent maintains the conversation.
2.     **ChEMBL:** The database which is used to retrieve information is ChEMBL. So, to access data from ChEMBL, ChEMBL web source client API is used[1]. This is the only official client Python library developed and supported by CHEMBL.
3.     **Telegram:** The bot is deployed on Telegram[2] which is a freeware, cross-platform and cloud-based instant messaging software and application service. It has been deployed using the Telethon library[3]. The interface is similar to a messaging application. User can select to perform an operation by choosing an operation from the button grid.
4.     **Cairosvg:** Images of the molecular structures of the drug molecules are displayed by the bot using the cairosvg library[4] which downloads an image from the URL and converts the SVG image to PNG format. To perform data manipulation and transformation operations, Pandas library has been used.

**Conversation with a bot can be categorised into two intents: casual talk and chemical talk.** In the casual talk, the bot is responsible for greeting the user and answering some basic questions about itself. In the chemical conversation, the bot can perform all the use cases explained in this section as per the user's requirement.

### 3.2. Operations

The bot starts when initiated to use with /start message. Then an API authentication process is completed. If the API key id is valid, a connection with the ChEMBL database is established. The Bot then displays an interface which allows users to select an option to carry out an operation. The user can interact with the bot by choosing the operation to perform from the options displayed. A grid of buttons is displayed which allows user to select one of the following options:

• Molecule Info: allows user to search for the information of the molecules.
• Tissue Info: allows user to get information about the target tissue.
• Similar compounds: allows user to search for similar compounds.
• Chat to Bot: allows user to chat with the bot.
• Exit: allows user to exit the operational interface.

The bot also allows user to download a csv file of the 50 approved drugs. The query is executed, and the result is displayed on the Telegram. The user can exit from the search interface by choosing the Exit option.

### 3.3. Workflow of Atreya

The workflow divided into two main components: conversational management and search management as shown in Figure 1. The details regarding the components are described as follow:

---

[1] https://github.com/chembl/chembl_webresource_client

[2] https://telegram.org/

[3] https://pypi.org/project/Telethon/

[4] https://pypi.org/project/CairoSVG/

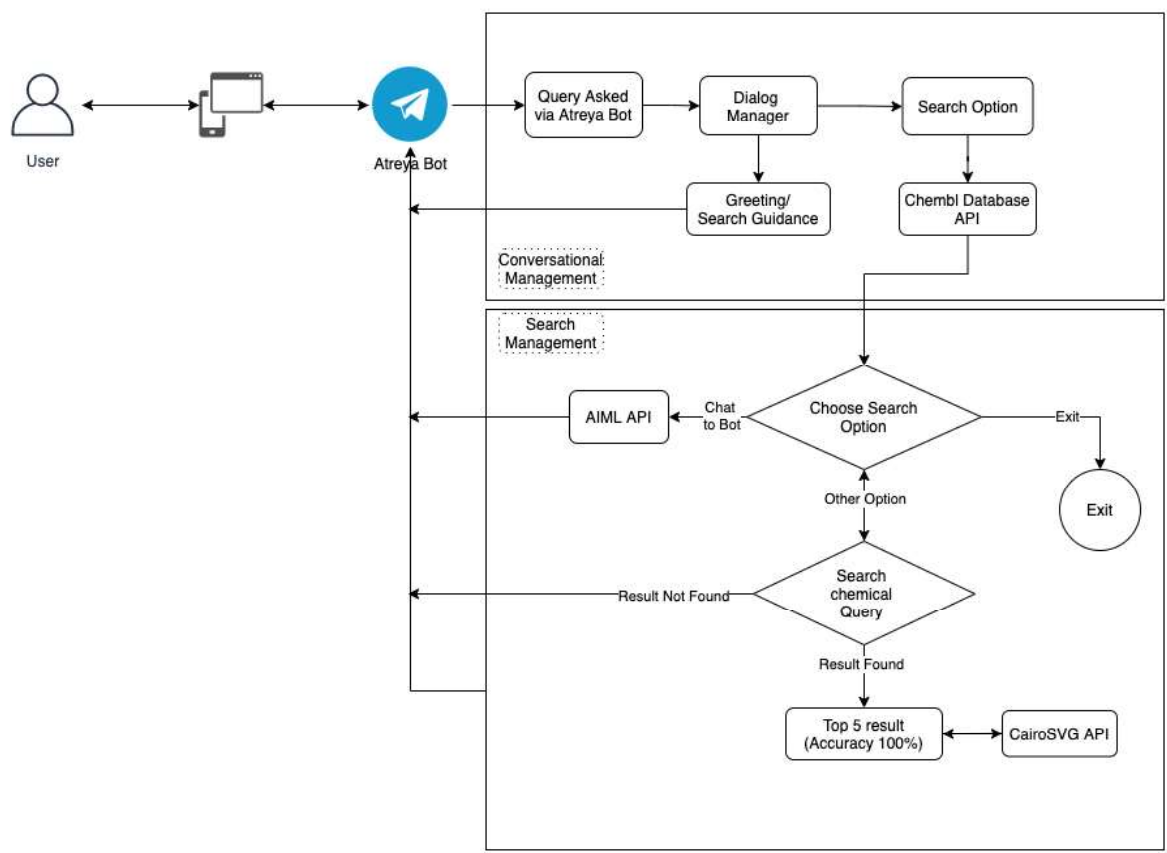

**Figure 1. Flow Diagram of Atreya Bot**

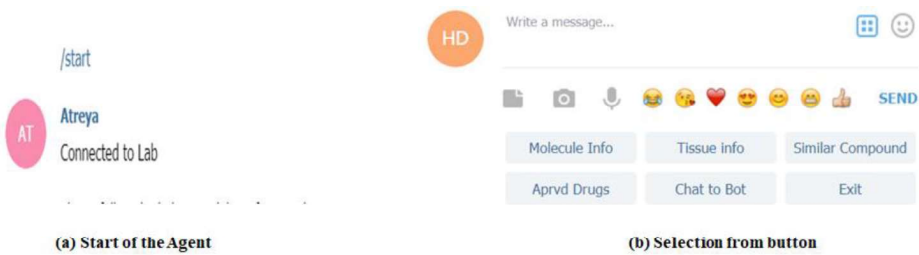

(a) Start of the Agent     (b) Selection from button

**Figure 2. Initial Steps of Atreya illustrating the start of the agent and button options**

### 3.3.1. Conversational Management

The bot allows user to perform an operation by either selecting an option from the menu displayed or by clicking on the buttons displayed. Figure 2 shows the initial steps that are taken in the Atreya bot. Users can also chat with the bot and search for their intended data. There is an option "Chat to bot" in the displayed options which allows users to implement the use cases discussed in the section 3.4. A use case will be selected by typing the specific keywords and then entering the desired query. For example, if the user wants to search for compounds similar to "Panadol", then after selecting the "Chat to bot" option, user will have to enter /msy:Panadol. Thus to initiate the search process, user will have to remember the keywords related to each use case. Please refer to this video demo[5] of the chatbot.

---

[5] https://drive.google.com/file/d/1LBM1a6RgfLp0n6ZG_nSuCvzAkrCI9GgM/view?usp=sharing

●     Dialog Manager: Dialog manager is used to manage the dialog structure. The input to the dialog manager is the user query and the search is performed with the help of search guidance. The subsequent dialog state is then provided to the user.
●     AIML-Bot-API: It is a simple restful API used to implement chatbots using AIML scripts.
●     ChEMBL web resource client: It is the official Python client library that can be used to access ChEMBL data and cheminformatics.

### 3.3.2. Search Management

This is responsible for searching and displaying the results. This bot is unique in the way that it displays the image of the molecular structure of the molecule user intends to search. Along with other details of the molecule, the bot also displays an image of the molecule which makes it more interactive. Another type of feature that is implemented in this bot is that it provides users to download a list of top 50 approved drugs in the csv format. Thus, by using a conversational interface, users can retrieve information from the ChEMBL database to suffice their requirements. The details as follows:

• Initiate the search process by starting the bot with command /start. The bot will display the list of use cases for the user to select from them using the keywords given aside every case.
• In the bottom widget, user can also select the button grid to display the menu from which an operation can be chosen.
• In order to search for information related to molecules, select the molecule info button.
• A list of guidelines will be provided from which the user can select the desired operation related to molecules.
• To search for a molecule name by Synonym, enter the keywords msy/Paracetamole (this is our desired search query in this example). Figure 3 illustrates a query for molecular information.
• On entering the query keywords, the bot will display the images of all the molecules synonymous to Paracetamol along with the ChEMBL Id, name and structure of the molecule displayed below the image of the molecule.

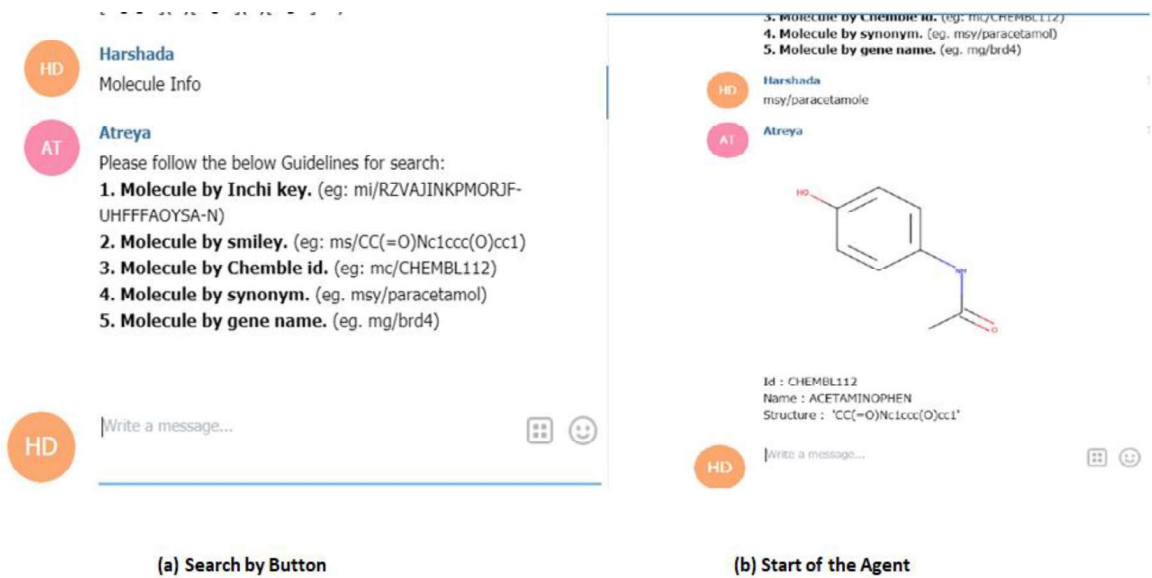

**Figure 3. Querying the molecular formulation**

### 3.4 Potential functionality and use cases
This section presents all the possible information-seeking service provided by the chatbot. Figure 4 subsequently represents the stages after the query is performed.

• **Search molecule name by Synonym:** This functionality will allow the user to search for a molecule by its specific name or by any synonym of the molecule which can be its common name too.
• **Search target by gene:** A drug target is basically a molecule, a protein that is associated intrinsically with a particular disease that can be addressed by a drug to produce the desired result. Atreya Bot allows the user to search for the target associated with a disease using the gene name.
• **Search compounds similar to the given drug:** User can search for compounds that are similar to the desired drug by entering the name of the drug. The Bot returns the SMILES of the similar compound, name and molecular formula. The similarity percentage is returned in descending order so that the user can select the most similar compound according to their relevancy.
• **Search molecule by SMILES:** Atreya bot provides users with a functionality to search for desired molecules by specifying the SMILES (simplified molecular-input line-entry system). The bot then returns the ChEMBL ID of the molecule with its molecular formula, InChi Key and name.
• **Search molecule by ChEMBL ID:** User can search for a molecule by mentioning the ChEMBL ID of the molecule. The bot returns the molecular formula, Inchi key and name of the molecule. Another unique functionality provided by the bot is that it shows the molecular structure of the molecule which will allow the user to get more details of the desired molecule.
• **Similar compounds by SMILES:** By specifying the SMILES, users can retrieve similar compounds from the database. The bot thus returns the name of the molecule, its ChEMBL ID and molecular structure.
• **Get tissue by Uberon ID:** Uberon is an anatomical ontology that represents body parts, organs and tissues in a variety of animal species, with a focus on vertebrates. The ontology includes comprehensive relationships to taxon-specific anatomical ontologies, allowing integration of functional, phenotype and expression data. Users can search for a tissue by specifying the UberonId.
• **Get tissue by name:** Users can search for a tissue by specifying the name.
• **Get the drug approval year by USAN stem:** United States Adopted Names (USAN) are unique non-proprietary names assigned to pharmaceuticals marketed in the United States. Users can get the approval year for a drug by specifying the USAN for the molecule. The bot returns classification code, first approval year, ChEMBL Id, canonical SMILES and synonym for the molecule.
• **Get approved drugs by disease name:** Users can get the approval year for a drug by specifying the name of the disease for which that drug is used. The bot returns first approval year, ChEMBL Id, name and formula for the molecule.
• **Get a list of all the approved drugs:** This unique functionality of the Bot returns a list of top 50 approved drugs in a csv format. The users who intend to get information about the drugs can download the file to serve their purpose.
• **Get tissue by ID:** Users can search for a tissue by specifying any Id such as BTO Id, Uberon Id, EFO ID, name, ChEMBL Id.

(a) Result after Query                                                                 (b) Results and Exits

**Figure 4. Results and Exits**

## 4.         CONCLUSION AND FUTURE WORK

This paper investigates the potential of the conversational search interface used by chemical scientists and chemistry students. This project's main objective is to produce an intelligent interactive IR System that reduces the user's efforts to find the relevant subject information. The project's future direction is to have chemistry teachers and students evaluate the bot using different usability metric types to explore user search experience and cognitive load.  The functionality of Atreya Bot is rule-based and simple in the paper. The simplicity was designed to provide a proof of concept and thus only simple queries were used. The next step in the project would be to incorporate more complexities into the system to account for factors such as previous knowledge of the user. A QA system can also be developed to support complex queries through the use of deep learning models.  A comparison of the Atreya Bot and state-of-the-art models can then be conducted to highlight the benefits and drawbacks.

# REFERENCES AND CITATIONS

Araujo, T. (2020). Conversational Agent Research Toolkit An alternative for creating and managing chatbots for experimental research1. Computational Communication Research, 2(1), 35–51. https://doi.org/10.5117/CCR2020.1.002.ARAU

Arguello, J., Choi, B., & Capra, R. (2018). Factors influencing users' information requests: Medium, target, and extra-topical dimension. ACM Transactions on Information Systems, 36(4). https://doi.org/10.1145/3209624

Avula, S., & Arguello, J. (2020). Wizard of Oz interface to study system initiative for conversational search. CHIIR 2020 - Proceedings of the 2020 Conference on Human Information Interaction and Retrieval, 447–451. https://doi.org/10.1145/3343413.3377941

Avula, S., Chadwick, G., Arguello, J., & Capra, R. (2018). SearchBot. 9–12. https://doi.org/10.1145/3272973.3272990

Davies, M., Nowotka, M., Papadatos, G., Dedman, N., Gaulton, A., Atkinson, F., Bellis, L., & Overington, J. P. (2015). ChEMBL web services: Streamlining access to drug discovery data and utilities. Nucleic Acids Research, 43(W1), W612–W620. https://doi.org/10.1093/nar/gkv352

Gwizdka, J. (2010). Distribution of cognitive load in Web search. Journal of the American Society for Information Science and Technology, 61(11), 2167–2187. https://doi.org/10.1002/asi.21385

Kaushik, A. (2019). Dialogue-based information retrieval. Lecture Notes in Computer Science (Including Subseries Lecture Notes in Artificial Intelligence and Lecture Notes in Bioinformatics), 11438 LNCS, 364–368. https://doi.org/10.1007/978-3-030-15719-7_50

Kaushik, A., & Jones, G. J. F. (2021). A Conceptual Framework for Implicit Evaluation of Conversational Search Interfaces. 1–14. http://arxiv.org/abs/2104.03940

Kaushik, A., Ramachandra, V. B., & Jones, G. J. F. (2020a). DCU at the FIRE 2020 retrieval from conversational dialogues (RCD) task. CEUR Workshop Proceedings, 2826, 788–805.

Kaushik, A., Ramachandra, V. B., & Jones, G. J. F. (2020b). An interface for agent supported conversational search. CHIIR 2020 - Proceedings of the 2020 Conference on Human Information Interaction and Retrieval, 452–456. https://doi.org/10.1145/3343413.3377942

Mendez, D., Gaulton, A., Bento, A. P., Chambers, J., De Veij, M., Félix, E., Magariños, M. P., Mosquera, J. F., Mutowo, P., Nowotka, M., Gordillo-Marañón, M., Hunter, F., Junco, L., Mugumbate, G., Rodriguez-Lopez, M., Atkinson, F., Bosc, N., Radoux, C. J., Segura-Cabrera, A., … Leach, A. R. (2019). ChEMBL: Towards direct deposition of bioassay data. Nucleic Acids Research, 47(D1), D930–D940. https://doi.org/10.1093/nar/gky1075

Radlinski, F., & Craswell, N. (2017). A theoretical framework for conversational search. CHIIR 2017 - Proceedings of the 2017 Conference Human Information Interaction and Retrieval, 117–126. https://doi.org/10.1145/3020165.3020183